# Orbital textures and evolution of correlated insulating state in monolayer 1T phase transition metal dichalcogenides

Qiang Gao[1]*, Haiyang Chen[1]*, Wen-shin Lu[2,3], Yang-hao Chan[2,3], Zhenhua Chen[4], Yaobo Huang[4], Zhengtai Liu[4], Peng Chen[1#]

[1]Key Laboratory of Artificial Structures and Quantum Control (Ministry of Education), Tsung-Dao Lee Institute, Shanghai Center for Complex Physics, School of Physics and Astronomy, Shanghai Jiao Tong University, Shanghai 200240, China.

[2]Institute of Atomic and Molecular Sciences, Academia Sinica, Taipei 10617, Taiwan.

[3]Physics Division, National Center for Theoretical Sciences, Taipei 10617, Taiwan.

[4]Shanghai Synchrotron Radiation Facility, Shanghai Advanced Research Institute, Chinese Academy of Sciences, 201204 Shanghai, China.

* These authors contributed equally to this work.

Email: pchen229@sjtu.edu.cn

Strong electron-electron interaction can induce Mott insulating state, which is believed to host unusual correlated phenomena such as quantum spin liquid when quantum fluctuation dominates and unconventional superconductivity through doping[1-7]. Transition metal compounds as correlated materials provide a versatile platform to engineer the Mott insulating state. Previous studies mostly focused on the controlling of the repulsive interaction and bandwidth of the electrons by gating or doping[8-10]. Here, we performed angle-resolved photoemission spectroscopy (ARPES) on monolayer 1T phase $NbSe_2$, $TaSe_2$, and $TaS_2$ and directly observed their band structures with characteristic lower Hubbard **bands. By systematically investigating the orbital textures and temperature dependence of the energy gap of the materials in this family, we discovered that hybridization of the chalcogen *p* states with lower Hubbard band stabilizes the Mott phase via tuning of the bandwidth, as shown by a significant increase of the transition temperature ($T_C$) at a stronger hybridization strength. Our findings reveal a mechanism for realizing a robust Mott insulating phase and establish monolayer 1T phase transition metal dichalcogenide family as a promising platform for exploring correlated electron problems.**

When repulsive Coulomb interaction is strong that conducting electrons becomes localized in system with half-filled bands, energy gap will be formed, leading to the Mott insulating ground state[1-3]. In a simplified Hubbard model, the existence of Mott phase is determined by the relative strength of Coulomb interaction ($U$) and the width of the half-filled band ($W$)[1]. Tuning of the interaction strength ($U/W$) can effectively drive the Mott transition in moiré superlattice via gating or in natural materials by doping[8-11]. Transition metal compounds with unpaired *d* electrons in transition metal ions can exhibit Mott insulating states[1,2], which is characterized by an opening of energy gap and emergence of flat bands known as Hubbard bands close to the Fermi level in the low temperature. Understanding of the Hubbard band structure is thus essential to the Mott physics.

$NbSe_2$, $TaSe_2$, and $TaS_2$ are members of the broad class of 1T phase layered triangular lattice materials. A first-order phase transition to a commensurate charge density wave (CDW) state has been reported in this family of materials[12-16]. As illustrated in Fig. 1a, the 1T phase $MX_2$ (M = Ta, Nb; X = Se, S) family has the same trigonal crystal symmetry but slightly different lattice constants. In a simple ionic picture, the chalcogen atoms form a distorted octahedron around the transition metal atom. The atomic *d* levels split and the energy of $d_{3z^2-r^2}$ orbital is lowered. The nominal $d^1$ valence electron results in half-filled $d_{3z^2-r^2}$ state and leads to a metallic band structure. In the low temperature, the in-plane atoms deformed into Star-of-David (SOD) clusters with a $(\sqrt{13} \times \sqrt{13})$ periodicity[17] (Fig. 1a). In a SOD unit cell, twelve in-plane Ta/Nb atoms distorted towards the central atom and twelve electrons with one from each transition metal atom occupy six full-filled bands. One unpaired electron localized on the central Ta atom forms a half-filled band which is predicted to cross the Fermi level from the conventional band theory. Previous experiments have shown that monolayer 1T $MX_2$ (M = Ta, Nb; X = Se, S) systems are insulating in commensurate CDW state with Star of David superstructure[18-22]. Mott-Hubbard mechanism was used to explain the ground state, but the Hubbard band fine structure has not been revealed and the effect of coupling with chalcogen orbitals remains elusive.

Here, we study the electronic structure of monolayer 1T-NbSe₂, 1T-TaSe₂, and 1T-TaS₂ by ARPES and show evolution of the Mott state with varying hybridization strength between ligand and transition metal states. Low dimensionality simplifies the interlayer interaction problem from various stacking orders[23-24] and reduces the screening around the Fermi level, which in principle increases the Coulomb interaction and leads to a more robust Mott state compared to the bulk counterpart[25-28]. Sharp diffraction patterns measured by reflection high energy electron diffraction (RHEED) indicates a well-ordered monolayer NbSe₂ grown by molecular beam epitaxy (MBE), as shown in Fig. 1c. ARPES spectra of monolayer NbSe₂ along $\overline{\Gamma M}$ and $\overline{\Gamma K}$ direction are shown in Fig. 1e. A nearly flat band at ~0.22 eV around the $\overline{\Gamma}$ point is observed[21,22,29,30]. The hole-like Se 4$p$ bands penetrate the flat band and exhibit a continuum structure with an inverted triangle shape. The diminished spectral intensity between the flat band and the Fermi level indicates a gap opening and the system is insulating in the low temperature. Note that the bands crossing the Fermi level at ~±0.5 Å$^{-1}$ are from the 1H-NbSe₂ as the 1T and 1H phase monolayer NbSe₂ coexist in the films. These metallic bands can be used as a reference to roughly confirm the position of the Fermi level. Dispersive Nb $4d$ band are evidenced at a larger momentum. No obvious change in these bands except the intensity with different photon energies (matrix element effect), in agreement with the 2D nature of the electronic structure of the system (Extended Data Fig. 3 and Extended Data Fig. 4).

High energy-resolution ARPES spectra of 1T phase monolayer $NbSe_2$, $TaSe_2$, and $TaS_2$ are shown in Fig. 2. The overall electronic structure including the band gap and flat band structures are similar in these materials and in excellent agreement with the calculated unfolded band dispersions (Fig. 2c). A ferromagnetic ground state and Hubbard U are included in the calculations in Fig. 2c to reproduce the experimental results. The location of lower Hubbard band (LHB) is determined by the $d_{3z^2-r^2}$ orbital projection from the spectra weight of the central transition metal atom in the $\sqrt{13} \times \sqrt{13}$ superstructure. The difference between the band structure of 1T phase $NbSe_2$ and $TaSe_2$ is mainly quantitative that Se bands are closer to the Fermi level so that there is a clear overlap (~37 meV) between the top Se $4p$ band and the LHB in monolayer $NbSe_2$. Such an overlap is absent in $TaS_2$ as the S $3p$ bands are in higher binding energies. In the $\sqrt{13} \times \sqrt{13}$ superstructure unfolded results (Fig. 2c), the LHB of these materials exhibit a dip structure around the $\bar{\Gamma}$ point with the maximum depth observed in $TaS_2$. ARPES spectra of monolayer $TaS_2$ exhibit a nearly dispersionless flat band, which is rationalized as the LHB fractionalized into a continuum structure because of the strong quantum fluctuations[4,31]. In $NbSe_2$ and $TaSe_2$, the dip is still visible in ARPES results and the structure exhibits an inverted triangle shape as the Se $4p$ states hybridize with the LHB.

We investigate the insulating ground state of 1T phase monolayer 1T-$MX_2$ (M = Ta, Nb; X = Se, S) by performing temperature dependent scans. As shown in Fig. 3, the flat band and the CDW folded bands become blurry with increasing temperature in monolayer $NbSe_2$. They are almost indiscernible at 330 K, which makes it difficult to extract the correct energy gap by fitting to a phenomenological function[32]. We symmetrized the energy distribution curves (EDCs) at the $\overline{\Gamma}$ point with respect to the Fermi level to illustrate the gap formation (Fig. 3d). The symmetrized EDC at 10 K shows a U-shaped structure around the Fermi level, indicating opening of a gap. At higher temperatures, a V-shaped structure is developed, demonstrating there are some spectral intensities filling up the gap and it is diminishing with increasing temperature. We extract the gap value using the leading-edge midpoint (LEM) of the energy distribution curves (EDCs). The square of the energy gap as a function of temperature follows a functional form described by a semi-phenomenological mean-field equation (blue solid curve in Fig. 3e). As the CDW state and Mott phase occur together, the fit yields a CDW-Mott transition temperature of $T_C = 553 \pm 12$ K for monolayer 1T-NbSe$_2$. We performed the same types of measurements for other 1T phase TMDCs and obtained $T_C = 353$ K and 479 K for monolayer 1T-TaS$_2$ and 1T-TaSe$_2$[33]. The transition temperatures for monolayers are higher compared to the bulk counterparts[12-15], illustrating the robustness of the Mott gap in 2D 1T phase TMDC family. Interestingly, the Mott gap in monolayer 1T-NbSe$_2$ is the smallest among these three materials, but the derived transition temperature is the highest. As shown in Fig. 3e, the gap is reduced much slower with increasing temperature in 1T-NbSe$_2$ compared to the other two systems and it becomes the largest at 330 K in NbSe$_2$. This scenario can be understood by a tunable bandwidth of the Hubbard band induced by varying hybridization strength between the chalcogenide band and the LHB in these materials.

To investigate the effects of $d$-$p$ band hybridization on the Hubbard band during the formation of $\sqrt{13} \times \sqrt{13}$ superstructure, we first study the orbital character of the Hubbard band structure and how the band is changed across the phase transition based on the first-principles results. The projected spectral weight of the Ta/Nb atoms in the (1 x 1) structure shows that $d_{3z^2-r^2}$ orbital dominates the conduction band crossing over the Fermi level (Extended Data Fig. 8). Similar dip structures in this conduction band are shown around the $\bar{\Gamma}$ point (Extended Data Fig. 7), which evolve into the dip structures of the Hubbard band in the Mott phase (Extended Data Fig. 13), demonstrating the origin of the Hubbard band structure around the $\bar{\Gamma}$ point. As shown in Fig. 4a and Extended Data Fig. 7, the depth of the dip, which determines the bandwidth of the Hubbard band, changes from 124 meV in the (1 x 1) structure to ~10 meV in the $\sqrt{13} \times \sqrt{13}$ superstructure in monolayer 1T-NbSe$_2$, and the extracted bandwidth in monolayer 1T-TaS$_2$ in the $\sqrt{13} \times \sqrt{13}$ superstructure is ~28 meV, which is the largest among these three compounds. Experimentally, an analysis of integrated EDCs around the $\bar{\Gamma}$ point (±0.2 Å) indicates full width at half maximum (FWHM) in monolayer 1T-NbSe$_2$ is the smallest in these systems (Extended Data Fig. 9), in consistent with the narrowest bandwidth in monolayer 1T-NbSe$_2$ predicted in the calculations. Note that it is difficult to accurately extract the bandwidth of the LHB from ARPES spectra as the broadening of the flat band from the hybridization.

Hybridization strength can be quantitatively evaluated from the spectral weight proportion of the $p$ states in the LHB. The overall hybridization strength in monolayer 1T-NbSe$_2$ is higher compared to the other two materials and the maximum hybridization is obtained at the position where the overlap occurs near the $\bar{\Gamma}$ point, as indicated by a blue arrow in Fig. 2c. It is ~50% in monolayer 1T NbSe$_2$, 12% higher than the monolayer TaS$_2$ case. The calculated charge density distribution of the LHB shows orbital textures in the Mott phase and the relative size of the $p$ orbital is the largest in NbSe$_2$ among these three systems (Extended Data Fig. 10). As a simple approximation, we extract the value of Coulomb energy ($U$) from the spectral Mott gap and obtain the bandwidth ($W$) of the Hubbard band from the calculated bandwidth of the metallic band in the CDW phase. As shown in Fig. 4b, the value of interaction strength ($U/W$) increases from monolayer TaS$_2$ to NbSe$_2$ by ~2 times following the trend of $T_C$, an effect from the narrowing of the bandwidth that suppresses the inter-site hopping of the electrons in monolayer NbSe$_2$. We plot them against the *d-p* band hybridization strength, verifying their correspondence and indicating the *d-p* band hybridization as an effective controlling parameter to tune the bandwidth and lead to a more robust Mott insulating ground state. It is shown that similar band hybridization may have major effects on the electronic structure of high temperature cuprate superconductors and play an important role in controlling the transition temperature[34-36]. A further computational study on the band structure of monolayer 1T-NbSe$_2$ is carried out by imposing various Se atomic displacements shifted along z axis to decrease the hybridization strength of Nb and Se orbitals (Extended Data Fig. 11). The resulting flat metallic band becomes broader at a larger displacement. It can be understood that more Nb *d* orbitals are involved in the formation of Nb-Nb bonds as the Nb-Se bond becomes longer and results in an increased Nb-Nb inter-site hopping at the smaller hybridization strength.

Note that the phase transition is complicated as a CDW transition is also involved in these systems. We estimated the lattice distortion (represented by the displacement of the transition metal atoms) induced in the CDW phase. The value is obtained to be of 6% of the lattice constant for monolayer TaS$_2$ and 8% for both the monolayer TaSe$_2$ and NbSe$_2$. Therefore, the smaller value of the transition temperature in monolayer TaS$_2$ is possibly partly due to a weaker CDW instability and the difference between TaSe$_2$ and NbSe$_2$ is dominated by the varying hybridization strength. On the other hand, a weaker CDW instability in monolayer 1T-TaS$_2$ combined with the on-site Coulomb interaction is supposed to result in a smaller opening energy gap. However, the measured LEM gap is 11 meV larger in monolayer 1T-TaS$_2$ compared to monolayer 1T-NbSe$_2$, indicating a difficulty of opening a gap when there is a band hybridization as it costs more energy to hybridize the LHB with the chalcogen $p$ bands.

In summary, we provide ARPES measurements directly show the detailed LHB structure in monolayer materials of 1T-$MX_2$ (M = Ta, Nb; X = Se, S) family, together with the excellent agreement with our first-principles calculations, establishing the tunable band structure and thus $U/W$ in 1T transition metal dichalcogenides by hybridization between the Hubbard band and chalcogenide band. Furthermore, possible quantum spin liquid behavior is revealed by the reduced intensity of the LHB and the closing of the energy gaps from the surface doping by magnetic atoms (Extended Data Fig. 5 and Extended Data Fig. 6). Our study provides physical insights into the band coupling issue and demonstrates a novel way for exploration of metal-insulator transitions.

**References**

1. Imada, M., Fujimori, A. & Tokura, Y. Metal-insulator transitions. *Rev. Mod. Phys.* **70**, 1039-1263 (1998).

2. Brandow, B. H. Electronic structure of Mott insulators. *Adv. Phys.* **26**, 652, (1977).

3. Mott, N. F. & Peierls, R. Discussion of the paper by de Boer and Verwey. *Proc. Phys. Soc.* **49**, 72-73 (1937).

4. Law, K. T. & Lee, P. A. 1T-$TaS_2$ as a quantum spin liquid. *Proc. Natl. Acad. Sci.* **114**, 6996-7000 (2017).

5. Balents, L. Spin liquids in frustrated magnets. *Nature* **464**, 199-208 (2010).

6. Dagotto, E. Correlated electrons in high-temperature superconductors. *Rev. Mod. Phys.* **66**, 763-840 (1994).

7. Lee, P. A., Nagaosa, N. & Wen, X. G. Doping a Mott insulator: physics of high-temperature

superconductivity. *Rev. Mod. Phys.* **78**, 17-85 (2006).

8. Li, T. et al. Continuous Mott transition in semiconductor moire superlattices. *Nature* **597**, 350-354 (2021).

9. Chen, G. et al. Evidence of a gate-tunable Mott insulator in a trilayer graphene moiré superlattice, *Nat. Phys.* **15**, 237 (2019).

10. Fath, M. et al. Spatially inhomogeneous metal-insulator transition in doped manganites. *Science* **285**, 1540 (1999).

11. Chatzieleftheriou, M. et al. Mott quantum critical points at finite doping. *Phys. Rev. Lett.* **130**, 066401 (2023).

12. Gelabert, M., Lachicotte, R. & DiSalvo, F. Insulator-metal and structural phase phenomena in $K_xBa_{1-x}CoS_2$ (x < 0.07). *Chem. Mater.* **10**, 613-619 (1998).

13. Di Salvo, F. J., Wilson, J. A., Bagley, B. G., & Waszczak, J. V. Effects of doping on charge-density waves in layer compounds. *Phys. Rev. B* **12**, 2220-2236 (1975).

14. Di Salvo, F. J., Maines, R. G., Waszczak, J. V., & Schwall, R. E. Preparation and properties of 1T-TaSe$_2$. *Solid State Commun.* **14**, 497-501 (1974).

15. Fazekas, P. & Tosatti, E. Electrical, structural and magnetic properties of pure and doped 1T-TaS$_2$. *Philoso. Mag. B* **39**, 229-244 (1979).

16. Colonna, S. et al. Mott phase at the surface of 1T-TaSe$_2$ observed by scanning tunneling microscopy. *Phys. Rev. Lett.* **94**, 036405 (2005).

17. Perfetti, L. et al. Spectroscopic signatures of a bandwidth-controlled Mott transition at the surface of 1T-TaSe$_2$. *Phys. Rev. Lett.* **90**, 166401 (2003).

18. Vaňo, V. et al. Artificial heavy fermions in a van der Waals heterostructure. *Nature* **599**, 582-586 (2021).

19. Lin, H. et al. Scanning tunneling spectroscopic study of monolayer 1T-TaS$_2$ and 1T-TaSe$_2$. *Nano Res.* **13**, 133-137 (2020).

20. Chen, Y. et al. Strong correlations and orbital texture in single-layer 1T-TaSe$_2$ *Nat. Phys.* **16**, 218 (2020).

21. Nakata, Y. et al., Robust charge-density wave strengthened by electron correlations in monolayer 1T-TaSe$_2$ and 1T-NbSe$_2$. *Nature communications*, **12**, 5873 (2021).

22. Zhang, H. et al. Impenetrable barrier at the Metal-Mott insulator junction in polymorphic 1H and 1T-NbSe$_2$ lateral heterostructure. *J. Phys. Chem. Lett.* **13**, 10713 (2022).

23. Ritschel, T. et al. Orbital textures and charge density waves in transition metal dichalcogenides. *Nat. Phys.* **11**, 328-331 (2015).

24. Butler, C. J., Yoshida, M., Hanaguri, T. & Iwasa, Y. Mottness versus unit-cell doubling as the driver of the insulating state in 1T-TaS$_2$. *Nat. Commun.* **11**, 2477 (2020).

25. Qiu, D. Y., Da Jornada, F. H. & Louie, S. G. Screening and many-body effects in two-dimensional crystals: monolayer MoS$_2$. *Phys. Rev. B* **93**, 235435 (2016).

26. Halperin, B. I. & Rice, T. M. Possible anomalies at a semimetal-semiconductor transition. *Rev. Mod. Phys.* **40**, 755-766 (1968).

27. Wang, Y. D. et al. Band insulator to Mott insulator transition in 1T-TaS$_2$. *Nat. Commun.* **11**, 4215 (2020).

28. Wang, Y. et al. Dualistic insulator states in 1T-TaS2 crystals. *Nat. Commun.* **15**, 3425 (2024).

29. Calandra, M., Phonon-assisted magnetic Mott-insulating state in the charge density wave phase of single-layer 1 T-NbSe$_2$. *Phys. Rev. Lett.* **121**, 026401. (2018).

30. Jiang, T. et al. Two-dimensional charge density wave TaX$_2$ (X = S, Se, Te) from first principles. *Phys. Rev. B* **104**, 075147 (2021).

31. He, W. Y. et al. Spinon fermi surface in a cluster Mott insulator model on a triangular lattice and possible application to 1T-TaS$_2$. *Phys. Rev. Lett.* **121**, 046401 (2018).

32. Norman, M. R., Randeria, M., Ding, H. & Campuzano, J. C. Phenomenology of the low-energy spectral function in high-$T_C$ superconductors. *Phys. Rev. B* **57**, R11093-R11096 (1998).

33. Chen, H. et al. Spectroscopic Evidence for Possible Quantum Spin Liquid Behavior in a Two-Dimensional Mott Insulator. *Phys. Rev. Lett.* **134**, 066402 (2025).

34. Ohta, Y., Tohyama, T. & Maekawa, S. Apex oxygen and critical temperature in copper oxide superconductors: universal correlation with the stability of local singlets. *Phys. Rev. B* **43**, 2968–2982 (1991).

35. Slezak, J. A. et al. Imaging the impact on cuprate superconductivity of varying the interatomic distances within individual crystal unit cells. *Proc. Natl. Acad. Sci. USA* **105**, 3203–3208 (2008).

36. Weber, C., Haule, K. & Kotliar, G. Apical oxygens and correlation strength in electron- and hole-doped copper oxides. *Phys. Rev. B* **82**, 125107 (2010).

**Fig. 1 | Structure and electronic bands of monolayer NbSe₂ films. a,** Schematic of the triangular lattice and displacements of metal atoms in the Star-of-David CDW cluster. **b,** The $d$ orbitals splitting in the distorted octahedral crystal fields. **c,** A RHEED pattern of a monolayer NbSe₂ film. **d,** Unfolded band dispersion for monolayer 1T-NbSe₂ with the $(\sqrt{13} \times \sqrt{13})$ superstructure along the $\overline{M\Gamma K}$ direction. **e,** ARPES spectra taken along the $\overline{\Gamma M}$ and $\overline{\Gamma K}$ directions for monolayer NbSe₂ with 40 eV $p$-polarized light.

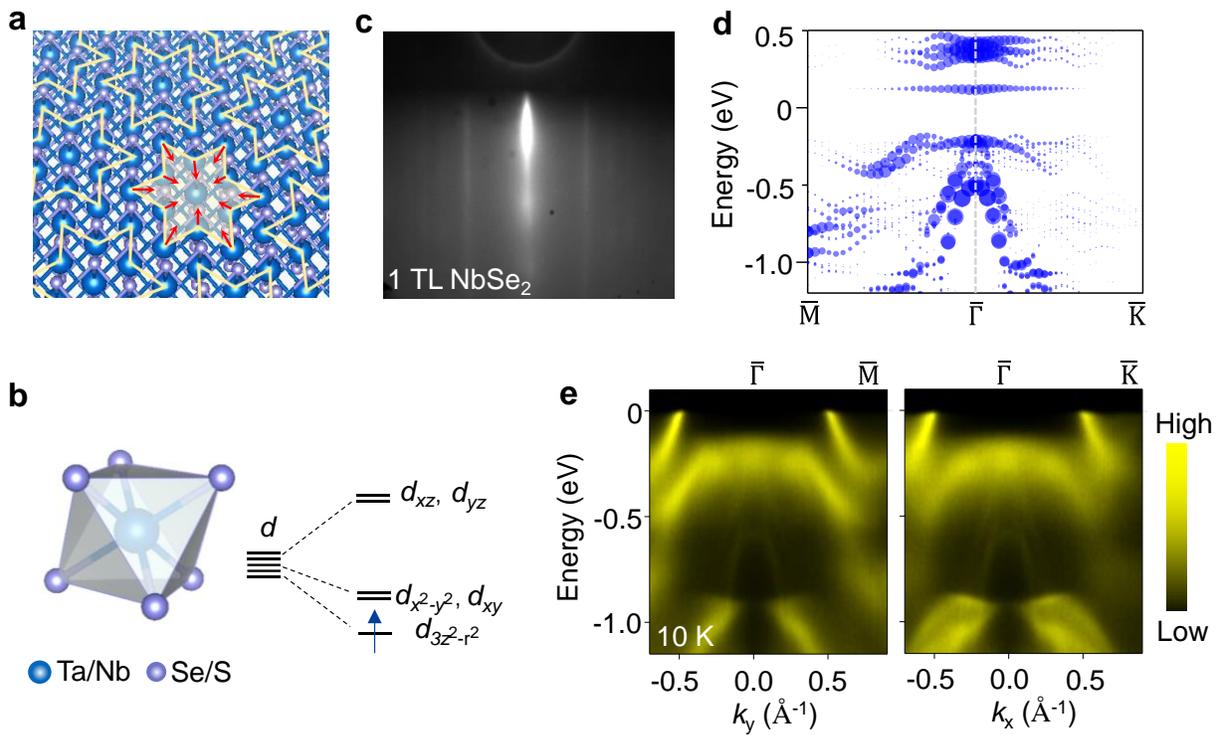

**Fig. 2 | ARPES spectra and orbital textures of monolayer NbSe$_2$, TaSe$_2$, and TaS$_2$ films. a**, ARPES maps taken along the $\overline{\Gamma M}$ direction with the photon energy of 40 eV at 10 K. **b**, Corresponding second derivative spectra for comparison. **c**, Calculated unfolded band dispersions for these three monolayer materials with the ($\sqrt{13} \times \sqrt{13}$) superstructure. The dispersions for the projected $d_{3z^2-r^2}$ orbitals (red dots) of the central metal atom are superimposed on the top. The magnitude of the spectral weight of the $d_{3z^2-r^2}$ orbitals is amplified by 4 times for clarity. The blue arrow indicates the position of overlap between the LHB and the $p$ state.

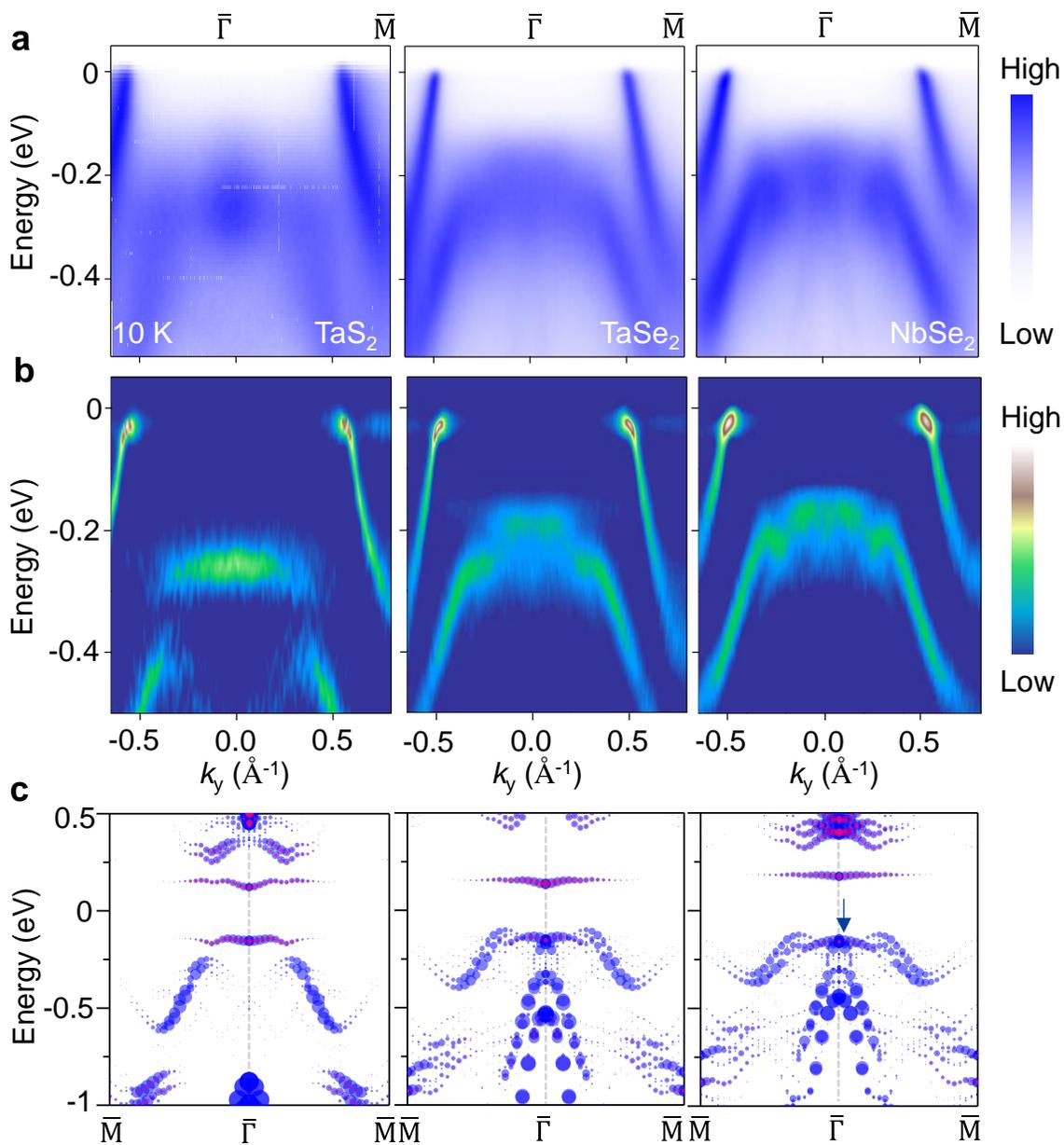

**Fig. 3 | Temperature dependence of the band structure and the Mott gaps. a,** Selected ARPES spectra of single-layer NbSe$_2$ taken along the $\overline{\Gamma M}$ direction between 10 and 330 K. **b,** Corresponding ARPES maps symmetrized in energy about the Fermi level show a Mott gap which is gradually reduced with increasing temperature. **c,d,** Normalized EDCs (**c**) and unnormalized symmetrized EDCs (**d**) at the zone center at selected temperatures between 10 and 330 K show the evolution of the gaps. The location of the leading-edge midpoints is indicated by a blue arrow. **e,** The extracted temperature dependence of the square of the Mott gap[33]. The solid curves are fitting results using the BCS-type mean-field equation. The pink triangles, green dots, and blue squares represent the results from TaS$_2$, TaSe$_2$, and NbSe$_2$, respectively.

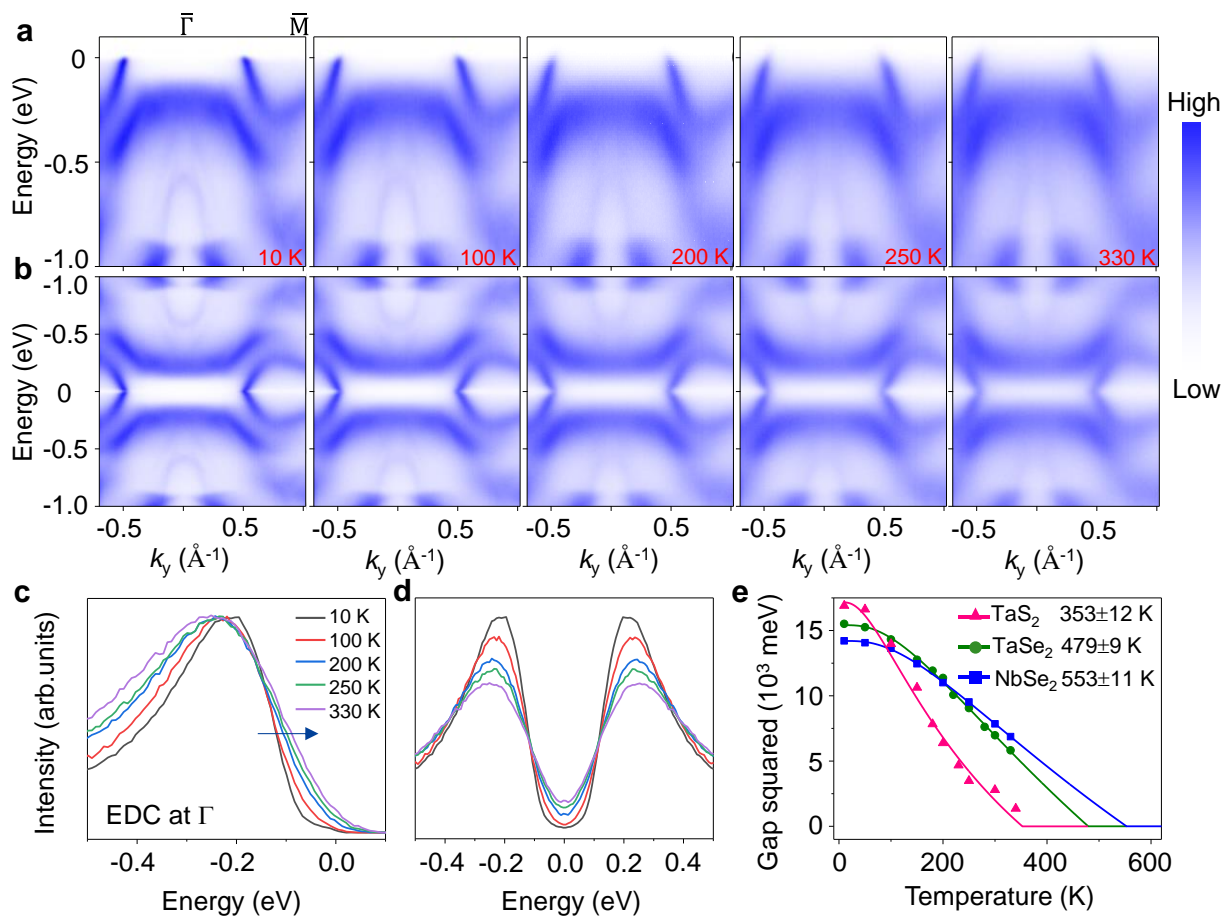

**Fig. 4 | Evolution of the band structure and transition temperature with *d-p* hybridization strength. a**, Calculated unfolded band dispersions for monolayer 1T-TaS$_2$, 1T-TaSe$_2$, and 1T-NbSe$_2$ with the ($\sqrt{13} \times \sqrt{13}$) nonmagnetic superstructure, showing the suppressed bandwidth of Hubbard *d* bands with increasing *d-p* hybridization. **b**, The extracted $T_C$ (red dots) for the three compounds and $U/W$ values (green inverted triangle) estimated from the Mott gap and bandwidth of metal band, plotted against the hybridization strength. Values of $U$ ($W$) are determined to be 130 (28), 125 (19), 119 (10) meV for monolayer 1T-TaS$_2$, 1T-TaSe$_2$, and 1T-NbSe$_2$, respectively. Note that $U$ might be underestimated as the spectral gap is determined from LHB to the Fermi level. Error bars are estimated from the uncertainty of the fitting.

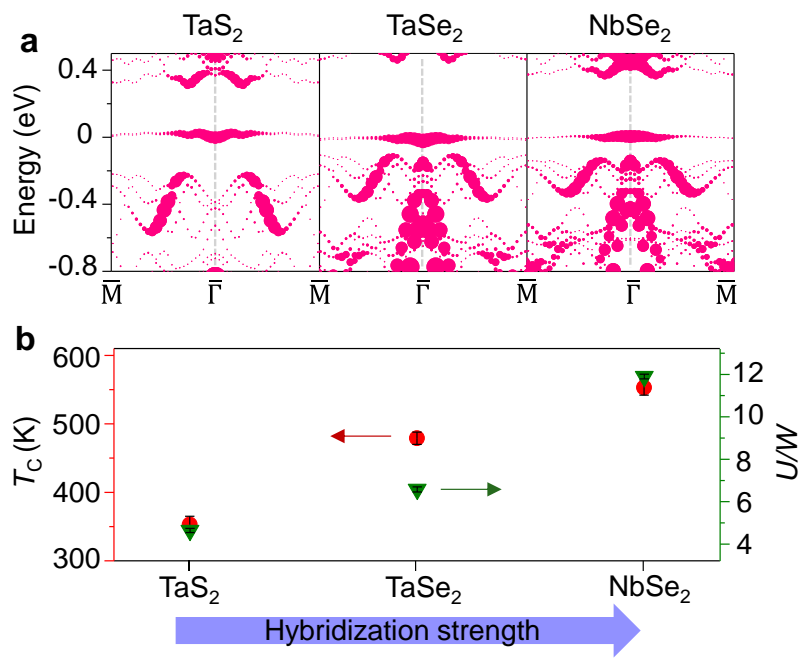

**Methods**

**Film growth**

Monolayer $NbSe_2$, $TaSe_2$, and $TaS_2$ thin films were grown on 4H-SiC substrates in the integrated MBE/ARPES systems at the lab in Shanghai Jiao Tong University. 4H-SiC were flash-annealed for multiple cycles to form a well-ordered bilayer graphene on the surface. The films were grown on top of the substrate by co-evaporating high purity Nb/Ta and Se/S from an electron-beam evaporator and a Knudsen effusion cell, respectively. The optimized substrate temperature was set at 600, 650, and 600 °C for $NbSe_2$, $TaSe_2$, and $TaS_2$, respectively. The growth process and thickness of the films was monitored by RHEED, and the growth rate was set to 30 minutes per layer of films. The formation of metastable 1T phase $TaSe_2$ is favored at higher substrate temperatures but with poor film quality and 1H phase dominates below 500 °C, as shown in Extended Data Fig. 1. Co atoms were evaporated from electron beam evaporators onto the samples in MBE at room temperature and then transferred into the ARPES system *in situ*. K atoms were deposited using outgassed SAES Getter sources onto the samples kept at 10 K in the ARPES system.

**ARPES measurements**

After growth, the films were transferred *in situ* to the ARPES system at the lab in Shanghai Jiao Tong University or transferred via a vacuum suitcase with a pressure better than $1 \times 10^{-10}$ mbar to the beamlines 03U and 09U at Shanghai Synchrotron Radiation Facility, in which the film surface was recovered through annealing at 300 °C before the experiments. ARPES measurements were performed at a base pressure of $\sim 5 \times 10^{-11}$ mbar with in-laboratory He discharge lamp (He-I 21.2 eV) and 30-100 eV photons at synchrotron using Scienta DA30 analyzers. Energy resolution is better than 15 meV and angular resolution is around 0.2°. Each sample's crystallographic orientation was precisely determined from the symmetry of constant-energy-contour ARPES maps. The Fermi level is determined by fitting ARPES spectra from a polycrystalline gold sample.

**Computational details from first principles**

The DFT calculations were performed using the Vienna ab initio package (VASP)[37,38] with the projector augmented wave method[39] and the Perdew-Burke-Ernzerhof (PBE) functional in the generalized gradient approximation (GGA)[40,41]. Spin-orbit couplings were self-included in the calculations. A plane-wave energy cut-off of 300 eV and a $7 \times 7 \times 1$ k-mesh were employed. Freestanding films were modeled with a 15-Å vacuum gap between adjacent layers in the supercell. Hubbard $U$ is included through the simplified rotationally invariant approach for each Ta or Nb atom to account for the electron localization. Spin-orbit couplings are included in the calculations. The band structure and the Mott gap depends on the choice of $U$ and lattice constants. An excellent agreement with the ARPES spectra is obtained with the in-plane lattice constants (Hubbard $U$) of 3.56 Å (2.5 eV), 3.53 Å (2 eV) and 3.36 Å (2.27 eV) for monolayer 1T phase $NbSe_2$, $TaSe_2$, and $TaS_2$, respectively. The atomic structure modulated by the SOD CDW phase was fully relaxed until the force on each atom was less than 0.005 eVÅ$^{-1}$. These parameters are consistent with those reported previously[22,27,30,42]. We used the VASPKIT code for band-unfolding calculations[43].

**Surface doping and quantum spin liquid behavior**

Mott insulator will exhibit quantum spin liquid behavior when the quantum fluctuations are strong enough to suppress the spin ordering[4,5,31]. Such a state can be revealed by probing the response of magnetic impurities doping[44,45]. We have employed in situ surface doping of Co on the monolayer films (Extended Data Fig. 5). Increasing amounts of Co doping causes the reduced intensity of the LHB and the closing of the CDW gaps in monolayer $TaSe_2$ at 10 K. This doping effect is more like a charge redistribution around the LHB and cannot be described by a rigid shift of the bands, as the Se bands shift further away from the Fermi level. The situation is more complicated for monolayer $NbSe_2$ as the gap is enlarged at the start of doping but reduced with an increased dosage, which suggests a small amount of Co atoms tends to occupy the off-center of the SOD. These results can be understood as a coupling between spin impurities and the chargons which gives rise to a charge redistribution and a reduced gap, a quantum spin liquid behavior in monolayer 1T $MX_2$ (M = Ta, Nb; X = Se, S). Instead, nonmagnetic impurities like K atoms shift the LHB bands away from the Fermi level and enlarge the energy gap for both monolayer $NbSe_2$ and $TaSe_2$ (Extended Data Fig. 6)[45,46].

**Data availability**

The data shown in the figures and other findings of this study are available from the corresponding authors upon reasonable request.

**Methods references**

37.  Hohenberg, P. & Kohn, W. Inhomogeneous electron gas. *Phys. Rev.* **136**, B864-B871 (1964).

38.  Kresse, G. & Furthmüller, J. Efficiency of ab-initio total energy calculations for metals and semiconductors using a plane-wave basis set. *Comp. Mater. Sci.* **6**, 15-50 (1996).

39.  Blöchl, P. E. Projector augmented-wave method. *Phys. Rev. B* **50**, 17953-17979 (1994).

40.  Perdew, J. P., Burke, K. & Ernzerhof, M. Generalized gradient approximation made simple. *Phys. Rev. Lett.* **77**, 3865-3868 (1996).

41. Langreth, D. C. & Mehl, M. J. Beyond the local-density approximation in calculations of ground-state electronic properties. *Phys. Rev. B* **28**, 1809-1834 (1983).

42. Darancet, P., Millis, A. J. & Marianetti C. A. Three-dimensional metallic and two-dimensional insulating behavior in octahedral tantalum dichalcogenides. *Phys. Rev. B* **90**, 045134 (2014).

43. Wang, V., Xu, N., Liu, J. C., Tang & G., Geng, W. T. VASPKIT: A User-Friendly Interface Facilitating High-Throughput Computing and Analysis Using VASP Code, *Computer Physics Communications* **267**, 108033 (2021).

44. He, W. Y. & Lee, P. A. Electronic density of states of a U(1) quantum spin liquid with spinon Fermi surface. I. Orbital magnetic field effects. *Phys. Rev. B* **107**, 195155 (2023).

45. Chen, Y. et al. Evidence for a spinon Kondo effect in cobalt atoms on single-layer 1T-TaSe$_2$. *Nat. Phys.* **18,** 1335-1340 (2022).

46. Lee, J., Jin, K. H. & Yeom, H. W. Distinguishing a Mott insulator from a trivial insulator with atomic adsorbates. *Phys. Rev. Lett.* **126**, 196405 (2021).

**Acknowledgments** We thank Prof. Yuanbo Zhang, Prof. Xiaoyan Xu, Dr. Xuejian Gao for helpful discussions. The work at Shanghai Jiao Tong University is supported by the Ministry of Science and Technology of China under Grant No. 2022YFA1402400 and No. 2021YFE0194100, the National Natural Science Foundation of China (Grant No. 12374188), the Science and Technology Commission of Shanghai Municipality under Grant No. 21JC1403000. Y.H.C. acknowledges support from the Ministry of Science and Technology, the National Center for Theoretical Sciences (Grant No. 110-2124-M-002-012), and the National Center for High-performance

Computing in Taiwan. Part of this research is supported by Shanghai Municipal Science and Technology Major Project. P. C. thanks the sponsorship from Yangyang Development Fund.

**Author contributions** P.C. conceived the project. Q.G., H.Y.C., P.C. with the aid of Z.H.C., Y.B.H., and Z.T.L. performed MBE growth, ARPES measurements, and data analysis. Y.H.C., and W.S.L. performed calculations. P.C., Q.G., and H.Y.C. interpreted the results. P.C. wrote the paper with input from other coauthors.

**Competing interests** Authors declare no competing interests.